\documentclass{article}

\usepackage{PRIMEarxiv}

\usepackage[utf8]{inputenc} 
\usepackage[T1]{fontenc}    
\usepackage{hyperref}       
\usepackage{url}            
\usepackage{booktabs}       
\usepackage{amsfonts}       
\usepackage{nicefrac}       
\usepackage{microtype}      
\usepackage{lipsum}
\usepackage{fancyhdr}       
\usepackage{graphicx}       
\graphicspath{{media/}}     

\usepackage{paralist}
\usepackage{hyperref}
\hypersetup{
colorlinks = true, 
linkcolor=black, 
citecolor=black, 
urlcolor=black 
}

\title{Foundation Models for the Digital Twin Creation of Cyber-Physical Systems
\thanks{All authors contributed equally. } 
}

\author{
  Shaukat Ali \\
  Simula Research Laboratory and \\ Oslo Metropolitan University \\
  Oslo, Norway \\
  \texttt{shaukat@simula.no} \\
   \And
  Paolo Arcaini \\
  National Institute of Informatics \\
  Tokyo, Japan \\
  \texttt{arcaini@nii.ac.jp} \\
  \And
  Aitor Arrieta \\
  Mondragon University \\
  Mondragon, Spain \\
  \texttt{aarrieta@mondragon.edu} \\
}

\begin{document}
\maketitle

\begin{abstract}
Foundation models are trained on a large amount of data to learn generic patterns. Consequently, these models can be used and fine-tuned for various purposes. Naturally, studying such models' use in the context of digital twins for cyber-physical systems (CPSs) is a relevant area of investigation. To this end, we provide perspectives on various aspects within the context of developing digital twins for CPSs, where foundation models can be used to increase the efficiency of creating digital twins, improve the effectiveness of the capabilities they provide, and used as specialized fine-tuned foundation models acting as digital twins themselves. We also discuss challenges in using foundation models in a more generic context. We use the case of an autonomous driving system as a representative CPS to give examples. Finally, we provide discussions and open research directions that we believe are valuable for the digital twin community.
\end{abstract}


\section{Introduction}\label{sec:introduction}
An increasing number of foundation models are being released, trained on vast amounts of data~\cite{FMIntro}. Some models are trained for a specific modality (e.g., text, audio, video), whereas others are trained on multi-modal data (e.g., a mix of text, audio, and video). Given that these models have learned generic patterns, there is an increasing interest to tailor these models for specific application contexts~\cite{FMOppor}, such as code~\cite{liu2024your} or model~\cite{ferrari2024model} generation. 

Digital twins (i.e., virtual replica~\cite{DTDef}) for Cyber-Physical Systems (CPSs) have been built with various methods, such as with the application of model-based systems engineering approaches~\cite{MDEDT} and machine learning techniques~\cite{DTSurvey}. In both cases, significant manual effort is required to create digital twins. With the emergence of powerful foundation models, it is natural to investigate whether such models can reduce the manual effort and the reliance on human expertise required to create digital twins. Such an investigation is the main objective of this paper. 

Building on our previous conceptual work~\cite{RE4DT,dtForCpsISOLA2020} focusing on digital twins for CPSs, we provide different perspectives on using foundation models for creating digital twin models and the capabilities they provide. In particular, we discuss foundation models used from two perspectives. First, we discuss various possibilities for using foundation models to generate digital twin models (e.g., in SysML and Open Modelica) and capabilities of digital twins (e.g., in simulation models or machine learning models). Second, we discuss the direct use of fine-tuned foundation models as the digital twin as a whole, which does not require any explicit digital twin model. In addition, we use autonomous driving as the CPS domain to illustrate possible solutions with fine-tuned foundation models. From each perspective, we provide detailed challenges and research opportunities that serve as a starting point to build a research agenda for using foundation models for CPSs, which can be extended to build digital twins for other systems (e.g., biological systems).

The rest of the paper is organized as follows: We start with relevant background in Section~\ref{sec:background}. Next, in Section~\ref{sec:contexts}, we provide the overall context of the use of foundation models in our case, followed by a discussion of the two perspectives in Section~\ref{sec:case1} and Section~\ref{sec:case3}. Finally, we provide discussions in Section~\ref{sec:discussion} and conclude the paper in Section~\ref{sec:conclusion}.

\section{Background} \label{sec:background}
In this section, we provide a brief background on foundation models in Section~\ref{subsec:fm}, followed by introducing the example case of developing digital twins for Autonomous Driving Systems (ADSs) in Section~\ref{subsec:example}.

\subsection{Foundation Models} \label{subsec:fm}
Foundation models have recently grabbed the attention of scientists, industries of all scales, public sectors, and even non-technical people such as artists~\cite{FMNewParadigm}. Such a model is a powerful machine learning model trained on a large amount of data, which has learned generic patterns. Consequently, for specific applications, foundation models need to be adapted (e.g., with fine-tuning , retrieval augmented generation (RAG) or prompt engineering) to improve their performance for the applications. Since these models are general models, their potential applications are even unknown by their developers. As a result, several opportunities exist to investigate the potential applications of these foundation models across different domains. Several applications exist across different domains, such as code generation in software engineering, various types of content generation, summarization, and other computer vision applications~\cite{FMIntro,FMNewParadigm,FM4MedicalAI}.

Different classes of foundation models exist, e.g., depending on the data on which these are trained. For example, large-language models (LLMs) are trained on vast amounts of textual data. Their examples include GPT-4, BERT, and Llama. Similarly, large vision models are another type of foundation models for computer vision, including CLIP~\cite{CLIP} and LandingLens~\cite{LandingAI}. In addition, multimodal foundation models are getting more popular, which are trained on multiple data modalities such as text, images, and video~\cite{multimodalfms}. Examples include Fuyu-8B~\cite{Fuyu} and ALIGN~\cite{ALIGN}.

\subsection{Example Case: Autonomous Driving Systems (ADSs)} \label{subsec:example}
ADSs are key parts of self-driving cars responsible for understanding environmental conditions, processing these conditions, and controlling the cars. To achieve such functionalities, ADSs rely on various sensors (e.g., LiDAR) to obtain environmental conditions, such as road and weather. Once such data about environmental conditions are obtained, ADSs process them and accordingly devise a driving plan, followed by controlling the vehicle to implement the plan. Due to safety considerations, lots of research is ongoing on making ADSs dependable, e.g., with automated testing~\cite{ADSSurvey}. Digital twins are also being investigated for ADSs for various purposes, e.g., running advanced simulations and testing~\cite{SLR4DT3ADS}. During the development of ADSs and its corresponding digital twins, foundation models could potentially help address several challenges. Examples include generating realistic environmental conditions for simulation (e.g., weather conditions), extracting realistic test cases from documented accident reports, and assisting developers in building digital twin models. Addressing such challenges could lead to various benefits, such as developing realistic ADS digital twins, decreasing the development cost of digital twins and ADSs, and improving their dependability.

\section{Overview: Application Contexts of Foundation Models for CPS Digital Twins} \label{sec:contexts}

\begin{figure}[!tb]
\centering
\includegraphics[width=0.5\linewidth]{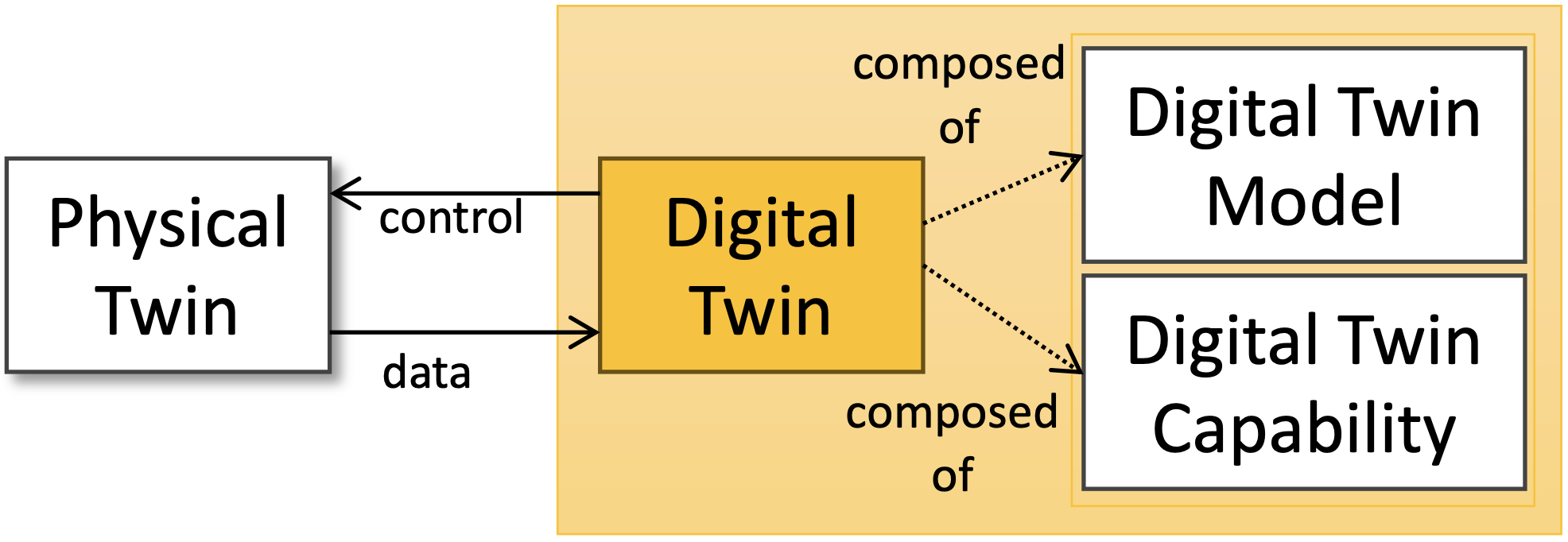}
\caption{Digital Twins for Cyber-Physical Systems}
\label{fig:dt}
\end{figure}

In our previous conceptual works~\cite{RE4DT,dtForCpsISOLA2020}, we defined the digital twin of a CPS (i.e., \textit{physical twin}) as consisting of two main components, i.e., the \textit{digital twin model} and the \textit{digital twin capability}, as shown in Figure~\ref{fig:dt}.
In this case, the model refers to a virtual replica or representation of the CPS, which simulates the CPS behavior. Such a model can be built with various approaches, such as Open Modelica~\cite{ModelicaDT}, SysML~\cite{Palachi2013}, or other simulation platforms, depending on the underlying design and development of the corresponding physical twin~\cite{Fitzgerald2019}. The digital twin capability refers to various functionalities of the digital twin, such as timely prediction of CPS anomalies before they actually happen~\cite{ATTAIN,SecurityDT2019,qinghuaDT}. A digital twin receives data from its corresponding physical twin, i.e., a CPS in our context. With the data, the digital twin can update its digital twin model and perform its intended functionalities facilitated by the digital twin model.

Based on the conceptual model described above, we identified two main cases in which foundation models can be used in the context of digital twins for CPSs to reduce their current purely manual development efforts. 

\section{Case 1: Foundation Models to Generate Digital Twins} \label{sec:case1}

We will discuss the first case, in which foundation models are used to generate digital twins (digital twin model, digital twin capability, or both) of a CPS. First, we will discuss state of the art related to this case, where AI techniques, in general, have been used to generate some aspect of digital twins of CPSs in Section~\ref{subsec:c1sota}. Next, in Section~\ref{subsec:c1proposal}, we will discuss a generic solution to generate digital twins of CPSs, followed by discussing challenges and opportunities related to generating digital twins with foundation models in Section~\ref{subsec:c1:challenges}. Finally, we present some examples using the case study of ADSs in Section~\ref{c1:casestudy}.

\subsection{State of the Art} \label{subsec:c1sota}
Since, in this case, we generate a digital twin model and a digital twin capability, we will discuss the state of the art related to these two aspects in the following.

Regarding generating a digital twin model, we used foundation models to generate such models, e.g., for stimulating a CPS. Depending on the CPS domain, such models could be developed using different formalisms, such as SysML and Open Modelica. To this end, manually creating digital twins with SysML models has been explored in the literature~\cite{SysMLDT1,SysMLDT2}. Similarly, Open Modelica has also been used to create digital twins of CPSs~\cite{ModelicaDT}. In addition, other model-driven approaches have been used to create digital twins of CPSs in various notations~\cite{MDEDT}. However, a study~\cite{GAIModeling} on using generative AI for modeling tasks revealed that the quality of current LLMs is quite poor for such modeling tasks. This suggests that generating digital twin models with foundation models is still an early area of research.

Concerning digital twin capability, various AI methods have been used to build various capabilities such as anomaly detection~\cite{ATTAIN,TCMSDT}, waiting time prediction for elevators' performance~\cite{ElevatorDT}, and parameter prediction in manufacturing~\cite{ParamPredict}. In our context, using foundation models to generate AI-based digital twin capability, including model structure, parameters, weights, etc., is a research area that needs further investigation. In addition, digital twin capabilities have also been built with traditional methods such as model-based simulation methods~\cite{CPSDTSurvey}. To this end, generative AI~\cite{GAIModeling} techniques need improvement and tailoring for this purpose, consequently requiring further investigation.

In summary, the current applications of generative AI techniques for developing models for digital twins are immature due to a lack of methods and tools to build specialized foundation models or fine-tune them to create digital twin models based on existing popular modeling languages for digital twins. Moreover, to generate AI-based capabilities, there is also a lack of methods and automated tools to create various model structures and their properties (e.g., weights).

\subsection{Solution Proposal} \label{subsec:c1proposal}
\begin{figure}[!tb]
\centering
\includegraphics[width=0.6\linewidth]{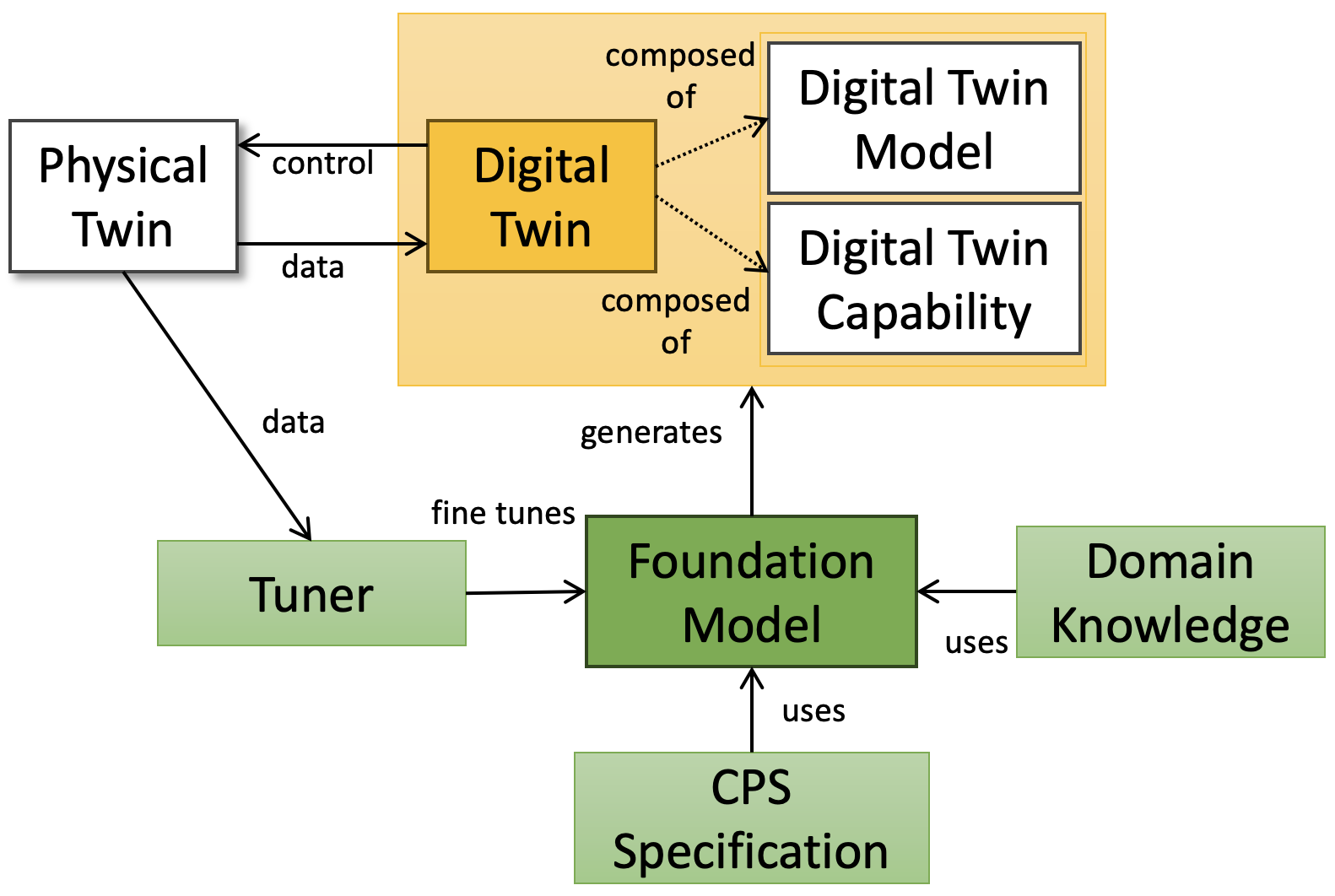}
\caption{Foundation models to generate a digital twin: (1) both the digital twin model and digital twin capability, (2) only digital twin model, (3) only digital twin capability}
\label{fig:c1}
\end{figure}

In this case (see Figure~\ref{fig:c1}), foundation models can be used to generate 1) the entire digital twin, i.e., the model and capability, 2) only the digital twin model, and 3) only the digital twin capability. 
Generally speaking, generic foundation models must be adapted to generate digital twin models and capabilities, e.g., fine-tuning parameters or prompt-tuning with the \textit{Tuner} component shown in Figure~\ref{fig:c1}. In addition, a \textit{Tuner} could be implemented with advanced machine learning techniques such as transfer learning. Moreover, to generate a more realistic digital twin, tailoring the foundation model will involve fine-tuning with CPS domain knowledge and the specifications of the specific CPS for which the digital twin is being generated. During the operation, data received from the real operation of CPS can be used further to update both the digital twin of the CPS and the foundation models with \textit{Tuner}. Below, we discuss the solutions concerning model generation and capability generation with foundation models.

\subsubsection{Digital Twin Model} \label{subsubsec:dtmodelgen}
When generating digital twin models with foundation models, we generate simulation models in various formalisms to assist in creating digital twin models.

Different degrees of automation are possible for this purpose, which can be supported by foundation models. 

\textbf{First}, foundation models can be used as a recommendation system, meaning that these models can help suggest different model elements when a user manually creates a simulation model. Various foundation models can be used depending on the type of CPS digital twin. For instance, if a simulation model is created in Open Modelica or SysML, LLMs are suitable, similar to their use in code generation. Depending on the formalism used, specific LLMs must be trained as a recommendation system as part of the modeling or simulation software. Other examples include creating a digital twin of the environment of a CPS, which could be a simple model, a digital shadow, or a combination of both. For example, in the case of a model, the environment could be represented as a video. To this end, video-based foundation models can recommend various new elements inside the digital twin. In the case of digital shadow, an initial model could be built with model-based approaches. Next, the model can continuously reflect the current state of the environment based on the data about the environment collected through the sensors.

\textbf{Second}, foundation models could be used as digital assistants in the form of chatbots to help create digital twin models. Compared to the recommendation system, this is a more interactive use of the foundation model, where a user interacts with the chatbot to create simulation models. The level of automation can vary. For instance, in simpler use, a user can ask for various suggestions, and the chatbot responds to them. In a more complex use, a user can even ask a chatbot to create model elements in a simulation model, the chatbot creates them, and the user confirms them.

\textbf{Third}, a more advanced use of foundation models would be to generate simulation models by providing the foundation models with a CPS specification, CPS models built during the design and development of CPS, and domain knowledge of CPS. Specifically, after an initial version of simulation models is generated classically, those can be further improved with interaction with foundation models.

\subsubsection{Digital Twin Capability}

Regarding building digital twin capabilities, we deal with generating a digital twin capability with foundation models.

To this end, foundation models could help in various ways, depending on how the capability is built. For instance, if the capability is built in a specific programming language, then quite a few specialized LLMs could be used. An interested reader is referred to the following reference for more details~\cite{LLM4CodeSurvey}. If the digital twin capability is built with simulation models similar to those described in the previous section, then a similar discussion applies here.

Increasingly, digital twin capability is also being built with machine learning models such as neural network models (e.g., \cite{sartaj2023hita,TCMSDT}). Foundation models could be used to create machine learning models to help create such machine learning models. LLMs could be used to create model structures, given that such models potentially have learned from many publicly available models. On the other hand, there is also the possibility that machine learning could be iteratively designed using LLMs as chatbots with whom a user interacts.

\subsection{Challenges and Opportunities} \label{subsec:c1:challenges}
Several challenges exist for the cases discussed in the previous section that provide research opportunities. Regarding generating digital twin models, there is not much work on generating models or assistance in general with various types of models that could be used for simulation. This is mainly because fewer open-source models (e.g., Simulink or Open Modelica) are available to train foundation models, e.g., compared to open-source code; thereby, more LLMs exist for generating the code. This would mean that research opportunities exist to adapt
foundation models at three levels. First, we must adapt foundation models for each formalism, such as Open Modelica, Simulink, and SysML. Second, such adapted foundation models must be adapted to certain domains based on domain knowledge, for instance, maritime and oil \& gas. At this level, tuning foundation models with domain-specific ontologies could help. Third, even after adapting such foundation models for each formalism and domain, such models further need to be fine-tuned for each specific CPS (e.g., with system specification or even existing CPS models that were created during the design of the CPS) for which the digital twin is being built. This is mainly because each CPS is built for a specific purpose, even within the same domain. Such tailored foundation models can support the various usages discussed in Section~\ref{subsec:c1proposal}, i.e., model element recommenders, modeling chatbots, and digital twin generators. 

In addition, we see some research opportunities for automated translation of digital twin models from one formalism to another with foundation models. This would help in cases where fewer models are available for training; thereby, models from various formalisms could be translated into one unified formalism and used to train foundation models for generating digital twins.

Regarding digital twin capability generation, once again, various opportunities exist. If the capability is built with simulation models, then the discussion related to digital twin models in this section is still applicable. Moreover, research opportunities exist to develop specialized chatbots that can assist a user in building machine-learning models-based digital twin capability.

\subsection{Example Case of ADS} \label{c1:casestudy}
Foundation models can be used from various perspectives in autonomous driving systems digital twins.

\textbf{First}, foundation models can automatically generate simulation models for the environment of the ADSs. We argue that foundation models have learned lots of public knowledge, e.g., about driving behaviors, accidents, weather conditions, human behaviors, and rules and regulations that can enable them to generate digital twins of the environment of the ADSs. Thus, such environmental digital twins could be integrated, for instance, with a particular autonomous driving algorithm and vehicle for the Software in the Loop (SiL) simulation. Nonetheless, this requires integration work with a particular ADS simulator being used. For example, environment digital twins can be used as time series data that can be input as part of an existing ADS simulator to generate a digital twin. For instance, LLMs could be relevant to this end. On the other hand, digital twins could be generated as videos that can be integrated into ADS simulators to mimic the environment. For such digital twins, foundation models that generate videos are relevant.

\textbf{Second}, foundation models can also support various digital twins' capabilities for ADSs built with classical methods. One such key functionality is to assess the realism of scenarios generated by digital twins, for example, to support the testing of ADSs as one of its key functionalities. To this end, a recent work~\cite{Wu2024RealityBA} evaluates the capability of commonly used LLMs, i.e., GPT, Llama, and Mistral, to assess the realism of scenarios generated by, e.g., reinforcement learning in the context of testing ADSs. Thus, such kind of realism assessment must be an integral part of digital twins of ADSs, either those digital twins generated by foundation models or only such assessment is implemented as part of digital twin implemented in other ways (e.g., with manual modeling and simulation approaches). Concerning this existing work~\cite{Wu2024RealityBA}, it only assessed the realism of textual scenarios. Therefore, several opportunities exist to determine the realism of scenarios generated as videos, images, or audio as part of digital twins of ADSs.

\section{Case 2: Fine-tuned Foundation Model as Digital Twin}\label{sec:case3}
\begin{figure}[!tb]
\centering
\includegraphics[width=0.6\linewidth]{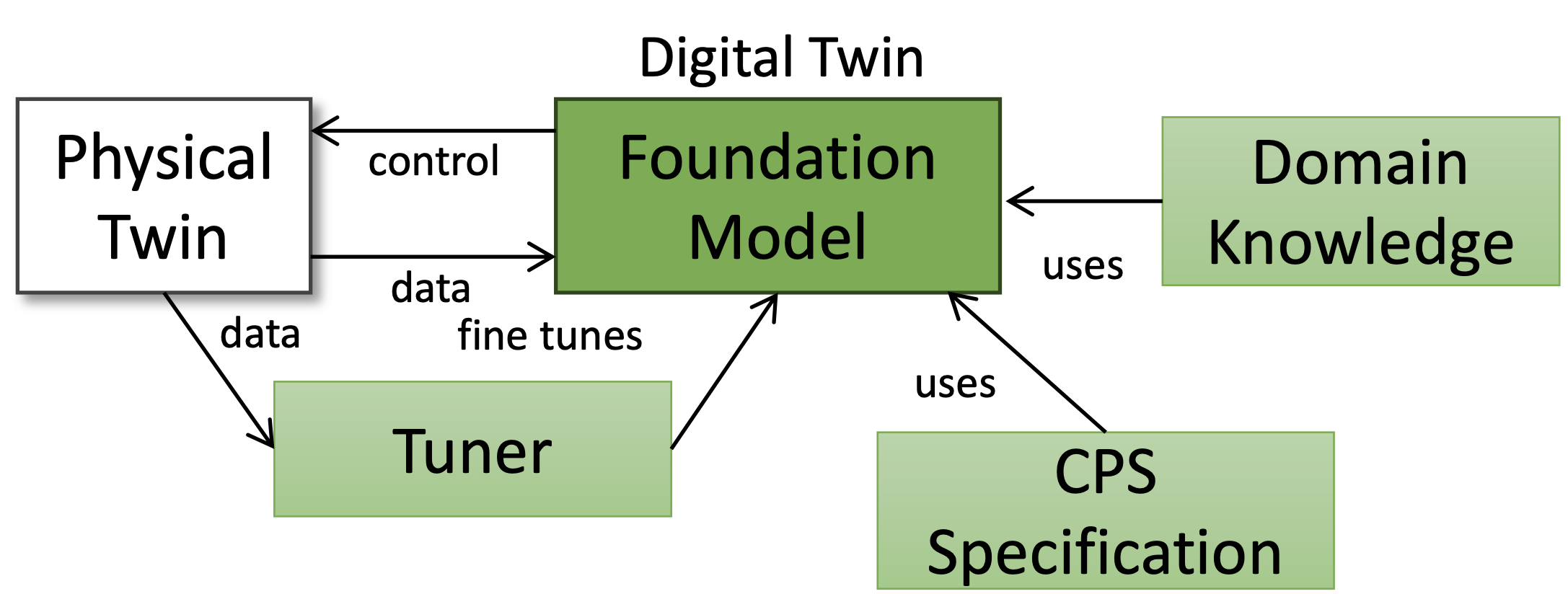}
\caption{Foundation model as a whole digital twin, as a digital twin model, or as digital twin capability}
\label{fig:case3}
\end{figure}

This is the advanced case, where a foundation model would serve as the backbone for the digital twin model only, the digital twin capability only, or the whole digital twin (see Figure~\ref{fig:case3}).
Such foundation model, i.e., the digital twin of the physical twin, can be for the digital twin model, digital twin capability, or the whole digital twin. To this end, foundation models must be fine-tuned with CPS specifications and other domain knowledge. Such fine-tuning can be done, for example, with prompt engineering or with advanced machine learning methods, e.g., transfer learning and few-shot learning. In addition, with more operational data coming from the physical twin, the digital twin (i.e., the foundation model) needs to be fine-tuned continuously.

\subsection{State of the Art} \label{subsec:c2sota}
Since Case 2 aims to construct a fine-tuned foundation model to play the role of the digital twin, in this section, we discuss the state of the art of developing fine-tuned foundation models for achieving relevant digital twin capabilities. 

A comprehensive survey of LLMs and their abilities is presented in~\cite{zhao2023survey}. The survey reports that the representative abilities of LLMs include \emph{language generation} such as code synthesis, \emph{knowledge utilization} such as knowledge completion, \emph{complex reasoning} such as symbolic reasoning, \emph{tool manipulation} such as search engine and interaction with external environment. The directly relevant one is the interaction with the external environment, where LLMs can obtain feedback from the external environment
and perform actions according to instructions in natural language, for instance, to manipulate agents. For example, VirtualHome, presented in~\cite{puig2018virtualhome}, is a virtual environment for simulating household tasks such as making coffee and cleaning, in which agents can execute natural language actions generated by LLMs. 

LLMs can also enable the creation of believable agents, which is useful for realizing certain digital twins that simulate human behaviors and their interactions with CPSs. For example, Part et al.~\cite{park2023generative} presented an agent architecture for generating believable behavior using LLMs. Moreover, Xi et al.~\cite{xi2023rise} presented a comprehensive survey on LLM-based agents and explored their applications in single-agent scenarios, multi-agent scenarios, and human-agent cooperation scenarios. 
Particularly, Yang et al.~\cite{yang2024llm} proposed an LLM-based digital twin to simulate human activities in the shopping mall to facilitate efficient training of a reinforcement learning algorithm to optimize user-preference-based temperature control. Moreover, Wang et al.~\cite{TWIN-GPT} proposed TWIN-GPT, which is fine-tuned on a pre-trained LLM on clinical trial datasets, and aims to generate personalized digital twins for different patients. 

LLMs can also be integrated with external knowledge bases or structured data sources to enhance their understanding and reasoning capabilities. Using external knowledge, an LLM-based digital twin can access domain-specific information by interacting with the physical twin. In the literature, some approaches have already been proposed to integrate LLMs with external information. For example, Jin et al.~\cite{Health-LLM} proposed to integrate health reports and medical knowledge into a large model to ask relevant questions for the prediction of diseases. 

In summary, there is a lack of methods and tools to develop virtual environments and agents with LLMs for constructing digital twins. Moreover, there is a lack of methods to exploit the LLM's interaction with the external environment and generate accurate actions, which could potentially enable the use of foundation models for developing digital twins that interact with their physical twin counterparts.

\subsection{Solution Proposal} \label{subsec:c2proposal}
\textbf{Foundation model being the digital twin model only.} 
In this scenario, a fine-tuned foundation model serves as the digital twin model while the digital twin capability is implemented with other ways. The fine-tuned foundation model is mainly about capturing the domain knowledge and understanding CPS behavior, while components implementing the digital twin capabilities handle the real-time interactions with the CPS. The capability components leverage the information provided by the fine-tuned foundation model (i.e., the digital twin model) to perform tasks such as monitoring, predicting, optimizing and controlling of the CPS. In this scenario, we can leverage the strengths of the foundation model for understanding and modeling complex systems. In addition, for digital twins that have high requirements on real-time interactions with the CPSs, decoupling the digital twin model from real-time processing provides the feasibility of allowing the fine-tuned foundation model perform computations offline or on a schedule, and process data in batches. Doing so allows for more efficient resource utilization, lower computation cost, and being more scalable.

\textbf{Foundation model being the digital twin capability only. }
In this scenario, a fine-tuned foundation model serves as the digital twin capability component, while the digital twin model is constructed as prior-knowledge based models (e.g., SysML models). This scenario is useful in application contexts where digital twin models are mandatorily required to provide a transparent view of the digital twin due to regulations, for instance. Moreover, with traditional digital twin models, fine-grained simulations and analyses can be performed, which empower engineers with more enhanced insights for risk assessment and management. Moreover, it might be possible that analysis and design models that have been already developed during the process of certain CPS development can be fully or partially reused as the digital twin models, which significantly reduces the cost of developing traditional digital twin models. 

The traditional digital twin model can also be used to derive data requirements and their dependencies which are useful information for guiding the selection of training data and preprocessing pipelines for the foundation-based digital twin models. In addition, traditional digital twin models may be used to generate prompts for LLM-enabled digital twin capabilities. For instance, if the digital twin model captures CPS anomalies, indicating places where the foundation model-based digital twin capabilities may need to be adjusted, then such anomalies can initiate fine-tuning processes for the foundation model. One can also benefit from the behavior simulation capability of the digital twin model by utilizing simulation outcomes as the indicator of generating prompts that reflect unexpected behavior, potential performance issues, etc.

\textbf{Foundation model being the whole digital twin.} In this scenario, the foundation model essentially is the whole digital twin and interacts with the CPS to provide real-time monitoring of the CPS, all kinds of analyses, and even the controlling of the CPS to various extents. One key advantage of this architecture is the fact that it significantly simplifies the development of the explicit digital twin model, which requires manual effort from domain experts who master at least one modeling notation and modeling tools such as OpenModelica and SysML, as we discussed in Case 1, where the digital twin model is an explicit representation of the physical twin. In addition, replacing the explicit digital twin model with the foundation model eliminates the large effort required to maintain the explicit digital twin model. This is because we envision that, with advanced AI technologies, the foundation model will self-maintain when receiving real-time operational data from the CPS. 

To fine-tune a foundation model-based digital twin, we must first choose a pre-trained foundation model. For instance, in autonomous driving, we can leverage large models to generate continuous video driving scenarios for testing ADS algorithms, detecting anomalies, identifying performance degradation, diagnosing issues via runtime monitoring, etc. Such a foundation model can already benefit from common-sense knowledge from pre-trained LLMs about the world, e.g., detecting traffic rule violations and anticipating potential safety threats. Which large models to choose as the base largely depends on the characteristics and requirements of the digital twin, the CPS and the end users' needs. For example, if the CPS involves processing a large amount of textual data, such as operating and maintenance logs, then LLMs might be suitable. Vision models might be more appropriate when involving videos and images. Also, vision-language models (VLMs) can process textual and visual information to enable multi-modal data processing. The field is rapidly booming, hence we envision that more and more large models will become available in the near future. 

After choosing a suitable foundation model and collecting available CPS data, documented domain knowledge, the digital twin, and CPS specification if it exists, the model can be fine-tuned to perform tasks. The process includes selecting model architecture, configuring the foundation model, selecting the fine-tuning strategy, training the fine-tuned model with the preprocessed CPS data, evaluating the model performance, and fine-tuning hyperparameters to obtain the best performance. Considering that the foundation model needs to be fine-tuned with real-time data, a tuner must be developed to streamline these fine-tuning tasks continuously and automatically.

\subsection{Challenges and Opportunities} \label{subsec:c2:challenges}
Though we see the clear benefits of fine-tuning a foundation model to act as a digital twin for a CPS of interest, several challenges must be overcome. 

First, selecting a suitable pretrained large model as the foundation model for developing the digital twin of a CPS is critical. This is, to a large extent, because it significantly impacts the performance and effectiveness of the digital twin. During the selection, it is important to systematically and comprehensively consider at least the following aspects: 1) the problem domain, including the CPS domain and its characteristics, types of data involved (e.g., sensor data, textual documentations), the complexity of tasks to be performed, and constraints on the deployment of the digital twin, etc.; 2) the availability and quality of data for fine-tuning, which is especially critical in the context of CPSs as we often do not have access to a large amount of distribution-wise balanced, high-quality data; and 3) clear understanding of the digital twin capabilities to be provided, such as anomaly detection for high-speed trains, safety threat prediction for autonomous driving, such that it makes it easier to decide which kinds of foundation models to choose. Currently, there are no guidelines or best practices for selecting suitable foundation models for developing digital twins for CPSs. Therefore, developing such guidelines is an important direction for future research.

Second, considering that the digital twin interacts with the physical twin at runtime, the efficiency of the digital twin is crucial. Especially in the case of the digital twin making decisions or providing crucial feedback to support the CPS operation in real-time, the digital twin has to efficiently process data, perform computations, and make decisions or generate responses. Therefore, pursuing a high-efficiency model operating as the digital twin is important, and advanced technologies will dramatically reshape how such a model is constructed and deployed in the future.

Third, considering the scenario that the whole digital twin is represented as a fine-tuned foundation model, we, therefore, do not have explicit representations of the digital twin. Hence, it is challenging to gain a comprehensive understanding of the digital twin's structure, behavior, and interactions with the CPS. Machine learning interpretability techniques (e.g., feature importance) might be needed to provide insights into the digital twin's behavior and enhance its transparency. However, such techniques are limited compared to the cases of explicit representation of the digital twin model. In addition, sufficiently verifying and validating the digital twin itself also becomes challenging. Alternative approaches, such as robustness testing and sensitivity analyses~\cite{UncerRobuaTOSEM}, might help assess the digital twin's performance and reliability without explicit representation of the digital twin model. For safety-critical CPS applications, building fine-tuned foundation models without having explicit representations might not be a viable solution, as the trustworthiness of the digital twins is paramount, especially when they directly interact with the safety-critical CPSs. All in all, we see challenges but also opportunities in this field. For example, established practices in the field of verifying and validating machine learning models can be adapted to this context. 

Fourth, when comparing digital twins with explicit representations to those based on fine-tuned foundation models, there are also challenges in fidelity that need to be considered. The former often captures detailed relationships, dependencies, and behaviors; however, fine-tuned foundation models typically abstract away details and hence reduce explicit modeling required complexity, therefore leading to a loss of fidelity in certain cases. How much fidelity is sacrificed largely depends on the CPS domain and expected digital twin capabilities. For instance, comparing an elevator system and autonomous driving, the loss of fidelity due to the lack of explicit representations should be larger for autonomous driving as it is much more complex, and hence, the abstraction gap is larger. Moreover, if an expected digital twin capability is trivial, we expect low demand for the digital twin fidelity. For instance, in the case of monitoring room temperature and humidity levels in a smart home, the expected digital twin capability is to provide alerts and notifications based on predefined thresholds. Therefore, the fidelity requirements for the digital twin are low. Consequently, the abstraction gap is small, and the impact of fidelity loss is minimal.

\subsection{Example Case of ADS} \label{c2:casestudy}
In the literature, there are proposals for utilizing LLMs for autonomous driving. For example, Cui et al.~\cite{cui2023large} proposed Talk2Drive, an LLM-based framework to process verbal commands from humans and make decisions autonomously based on contextual information. Talk2Drive is equipped with a speech recognition module for translating verbal inputs from humans to textual instructions, which are then sent to LLMs for reasoning and, afterwards, generating control commands, which are then automatically executed with the control module. Talk2Drive can be considered a personalized digital twin for autonomous driving, aiming to reduce the human driver takeover rate of autonomous vehicles. We also foresee that LLMs can be utilized to generate digital twins for monitoring the safety of autonomous driving and, when needed, triggering safety mechanisms (e.g., providing early warnings or recommending risk mitigation strategies) and assessing the severity of safety hazards, assisting in emergencies to ensure safe operation of ADSs. Moreover, foundation model-based digital twins can continuously learn and refine their capabilities over time.

\section{Discussions} \label{sec:discussion}
\subsection{Uncertainty of Foundation Models}
Though, as we discussed in the previous sections, foundation models offer opportunities to develop digital twins of various forms, their applications, especially in safety-critical and mission-critical domains, raise significant concerns about their inherent uncertainty. For example, Huang et al.~\cite{huang2023look} empirically explored different quantification methods of epistemic uncertainties of LLMs, aiming to understand whether and how well uncertainty estimation can help characterize LLMs' capabilities in performing different tasks. Tanneru et al.~\cite{tanneru2024quantifying} proposed novel methods for quantifying uncertainty in natural language explanations of LLMs. With such efforts on understanding and even quantifying uncertainties of large models, we foresee that there is still a long way to go to have reliable and trustworthy foundation models to apply. This is an active and ongoing research area on its own. For CPS domains that are highly concerned with safety, security, etc., cautions should be made to ensure the deployment of foundation models is safe and reliable, such as conducting thorough risk assessments and adhering to rigorous validation and verification processes.

\subsection{Cost and Effectiveness of Using Foundation Models}
For any software engineering solutions, we need to thoroughly consider their cost-effectiveness. Applying foundation models does not come with no cost. First, developing foundation models can involve high costs of computational resources, data collection and preprocessing, model training, fine-tuning, etc. Such costs can vary across different CPS domains. Second, deploying certain foundation models requires specialized hardware resources (e.g., GPU, FPGA), especially considering that fine-tuned foundation models (directly acting as digital twins -- Case 2) need to communicate with CPSs, which demands high throughput and low latency of communication and computation. In addition, fine-tuned foundation models must be maintained regularly, contributing to maintenance costs. However, as discussed in previous sections, there are advantages in applying foundation models, such as reducing the manual effort required to build traditional digital twin models. Therefore, in our opinion, it is necessary to compare the cost-effectiveness of using foundation models against traditional solutions via conducting empirical studies, such that using foundation models is justified. 

\subsection{Lightweight Foundation Models for Digital Twins} 
In addition to tailoring large foundation models for use as digital twins, we also foresee that there will be a need for lightweight specialized foundation models in some contexts. This will, in particular, be important in cases where digital twins are to be executed on dedicated hardware with low computational power. Moreover, we foresee their applications in critical applications where large foundation models cannot be accessed over the cloud. In addition, these models will be suitable for digital twins where inference time is important since the inference time of lightweight models is supposed to be shorter. Consequently, these foundation models are more suitable for digital twins with real-time applications. Finally, since these models can be deployed on dedicated resources, there will be less need to send data, e.g., to the cloud, thereby more security and privacy-preservation.

\subsection{Challenges in Generating Digital Twins with Foundation Models}
Using foundation models for the digital twin of CPSs comes with challenges, which generally hold for such models. For example, how to deal with hallucinations when generating digital twins or using them as digital twins. To this end, studying and dealing with {\it hallucinations} in foundation models (i.e., content generated by the foundation model that is not based on factual information) is an active research field with several existing strategies, and more strategies are appearing quickly~\cite{FMHallucinationSurvey,LLMHallucinationSurvey}. Investigation of existing hallucination handling strategies is an interesting research direction to follow.

Another challenge is how to assess the fidelity of digital twins. There are two aspects. First, in terms of using foundation models to generate digital twins, the fidelity of digital twins can be accessed similarly as typically done (e.g., see~\cite{dtForCpsISOLA2020}). Second, when foundation models are used as digital twins, it opens up a new research direction for assessing their fidelity, since this will require defining new metrics and methods.

Finally, foundation models pose many challenges related to ethical and legal aspects. Using digital twins to generate foundation models naturally leads to inheriting those challenges. For example, using foundation models to generate digital twins may generate copyrighted models. Moreover, these models are digital twins if such models have been trained on private data in certain cases. Furthermore, another challenge is whether digital twins will not make discriminatory and biased decisions is another challenge~\cite{FMOppor}.

\section{Conclusion}\label{sec:conclusion}
Due to their generic nature, foundation models offer many potential applications. To this end, software engineering researchers have increasingly been interested in discovering such applications. Along this line, we present perspectives on how foundation models can be used in creating digital twins of cyber-physical systems (CPSs). In particular, we present two perspectives, i.e., using foundation models to generate digital twins and using foundations as digital twins. We present state-of-the-art conceptual solutions, challenges, and opportunities for each perspective. We also use the case of digital twins of autonomous driving systems to present some ideas. Finally, we present discussion and open research questions.

\section*{Acknowledgments}
S. Ali is supported by the Co-tester project (No. 314544) funded by the Research Council of Norway and the RoboSAPIENS project financed by the European Commission’s Horizon Europe programme under Grant 101133807. P. Arcaini is supported by MIRAI Engineerable AI Project (No. JPMJMI20B8), JST. Aitor Arrieta is part of the Software and Systems Engineering research group of Mondragon Unibertsitatea (IT1519-22), supported by the Department of Education, Universities and Research of the Basque Country.

\bibliographystyle{unsrt}  
\bibliography{biblio}

\end{document}